\documentstyle[psfig,natbib,aas_macros]{gmn}

\def\newblock{\hskip .11em \@plus.33em \@minus.07em}%

\newif\ifAMStwofonts

\newcommand{\msun}{\mbox{${\rm M}_{\sun}$}}

\title[The binary nature of AM CVn] {Spectroscopic evidence for
the binary nature of AM CVn}

\author[Nelemans, Steeghs and Groot] 
       {G.~Nelemans$^1$, D.~Steeghs$^2$, P.~J.~Groot$^3$ \\
        $^1$ Astronomical Institute, University of Amsterdam, 
        Kruislaan 403, 1098 SJ, Amsterdam, The Netherlands {\tt (gijsn@astro.uva.nl)} \\ 
        $^2$ Astronomy Group, University of Southampton, Highfield,
        Southampton, SO17 1BJ, UK {\tt (ds@astro.soton.ac.uk)}\\ 
        $^3$ Harvard-Smithsonian Center for Astrophysics, 60 Garden
        St., Cambridge, MA 02138, USA {\tt (pgroot@cfa.harvard.edu)} }
\date{ Received \today}

\pagerange{\pageref{firstpage}--\pageref{lastpage}}

\pubyear{2000}

\begin{document}

\maketitle

\label{firstpage}

\begin{abstract}
  We analysed archival spectroscopic data of AM CVn taken with the
  William Herschel Telescope in 1996. In the literature two orbital
  periods for AM CVn are proposed. A clear S-wave in the He\,{\sc i}
  4471, 4387 and 4143~\AA\ lines is revealed when the spectra
  are folded on the 1029~s period. No signature of this S-wave is seen
  when folded on 1051~s. Doppler tomography of the line profiles shows
  a clear signature of the hotspot. Using this we can constrain the
  value of $K_2$ to lie between 210 and 280 km s$^{-1}$. Our work
  confirms the binary nature of AM CVn beyond any doubt, establishes
  1028.73~s as the true orbital period and supports the interpretation
  of AM CVn as a permanent superhump system.
\end{abstract}


\section{Introduction}\label{introduction}

AM CVn (HZ 29) was found as a faint blue object by \citet{hz47} and
shows broad He\,{\sc i} absorption lines \citep{gm57} and no hydrogen.
Periodic brightness variations on a period of $\sim$18 min.
\citep{sma67} and flickering \citep{wr72} suggested mass transfer in a
very compact binary. A model in which a degenerate helium white dwarf
transfers helium to another white dwarf, driven by the loss of angular
momentum due to gravitational wave radiation was proposed by
\citet{pac67} and \citet*{ffw72}. See for details on the models for AM
CVn stars and a list of the 8 currently known systems \citet{npv+00}.

Since the discovery of the periodic brightness variations and the
flickering, AM CVn has been extensively studied with high speed
photometry \citep[e.g][]{spk+99,spb+98} yielding multiple
periodicities on 1051, 1011 and 1029~s \citep[see
also][]{sea91,hsk+98}. Spectroscopic observations \citep*{phs93} show
a 13.38~hr periodicity in the skewness of the He\,{\sc i} absorption
lines.

According to \citet{spb+98} the dominant 1051~s photometric period is
the orbital period and the 1029~s a beat between the orbital period
and the 13.38~hr precession period of the accretion disc. However,
\citet{spk+99} argue that 1029~s is the orbital period and 1051~s is
the beat period. In the latter case, the system would be similar to
the permanent superhump systems among the hydrogen rich cataclysmic
variables \citep{spk+99}.

To discern between these two possibilities and to establish beyond
doubt the binary nature of AM CVn, a spectroscopic signature at either
period is needed. In this article we describe a spectroscopic study of
AM CVn in which we found a clear signature on the 1029~s period,
proving that this is the orbital period of the system.

\section{Data reduction}\label{reduction}

We analysed archival spectroscopic data obtained on the 4.2m William
Herschel Telescope on February 26 and 27, 1996, with the ISIS
spectrograph. In the first night a dichroic was used with both the
ISIS red and blue arm. Because of ripples in the blue part of the
spectrum introduced by the dichroic, it was removed on the second
night and only blue spectra were taken. Because of the ripples and the
lack of flat-fields of high quality for the first night we mainly
discuss the data taken on the second night. In the blue arm the
spectra were obtained with the R600B grating covering 3960-4760~\AA\ 
at a resolution of 2.0~\AA.

There are 403 spectra taken on the second night (173 on the first),
each with an integration time of 30~s.  Wavelength calibration lamp
exposures were taken approximately every 50 spectra (about every 40
minutes). The spectra were reduced using the standard data reduction
package MIDAS. After bias subtraction and flatfielding
the spectra were extracted and wavelength calibrated. The wavelength
calibration was very stable (less than half a pixel change between the
subsequent arc spectra) and the calibration of each arc spectrum was
used for the 25 spectra taken before and the 25 spectra taken after
the arc without interpolation. The flat-field for the first night was
taken at the beginning of the second night and had to be corrected for
a small wavelength shift.

\begin{figure}
\psfig{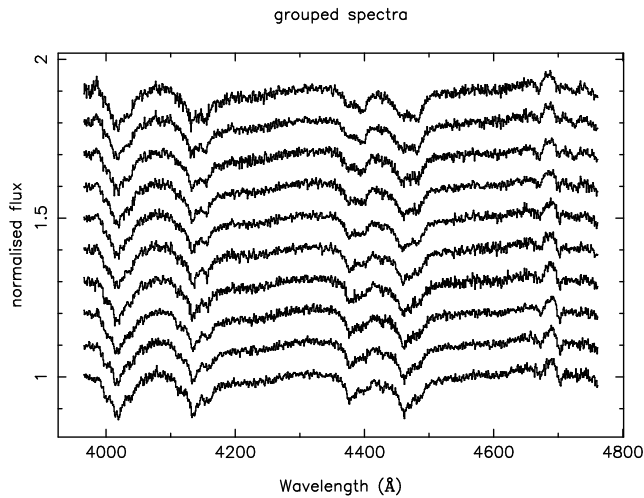}
\caption[]{
  Spectra of the second night averaged into 10 groups of 40
  consecutive spectra added together after normalising the continuum.
  The spectra show the change in the shape of the absorption lines
  over the 5 hours of the observation (see Sect.~\ref{precession}).}
\label{fig:spectra}
\end{figure}

The continuum of the wavelength calibrated spectra was fitted with a
cubic spline of order 3 to line-free areas and used to normalise all
spectra with respect to their continuum level. In
Fig.~\ref{fig:spectra} we show the resulting spectra of the second
night after averaging them into 10 groups over the whole night.

\section{Data analysis}\label{analysis}

\subsection{A period search}\label{period}

\begin{figure}
\psfig{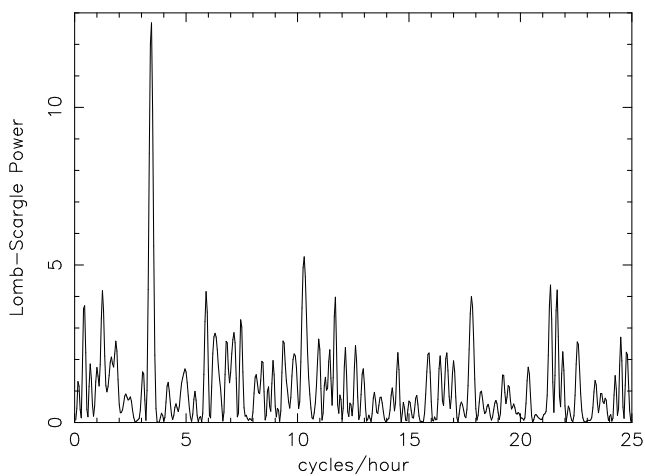}
\caption[]{Lomb-Scargle power spectrum of the equivalent width lightcurve
  of the 4 strongest He\,{\sc i} lines. The peak at 3.5 cycles/hour
  points to an orbital period around 1030s. }
\label{power}
\end{figure}

In order to search for line profile variations on the uncertain
orbital period of the binary, an equivalent width (EW) lightcurve was
constructed using the combined EW of the four strongest He\,{\sc i}
lines.  A Lomb-Scargle power spectrum (Fig.~\ref{power})  reveals a
clear peak around 3.5 cycles/hour, close to the expected orbital
period of 17 minutes.

Because the accuracy of the candidate orbital periods as derived from
the photometric studies of AM CVn exceeds the accuracy of our period
determination using the EW lightcurve, we folded the data set on the
two proposed orbital periods, 1028.73~s \citep{spk+99} and 1051.2~s
\citep{spb+98}.  Both periods lie very close to the peak in our power
spectrum.

\begin{figure}
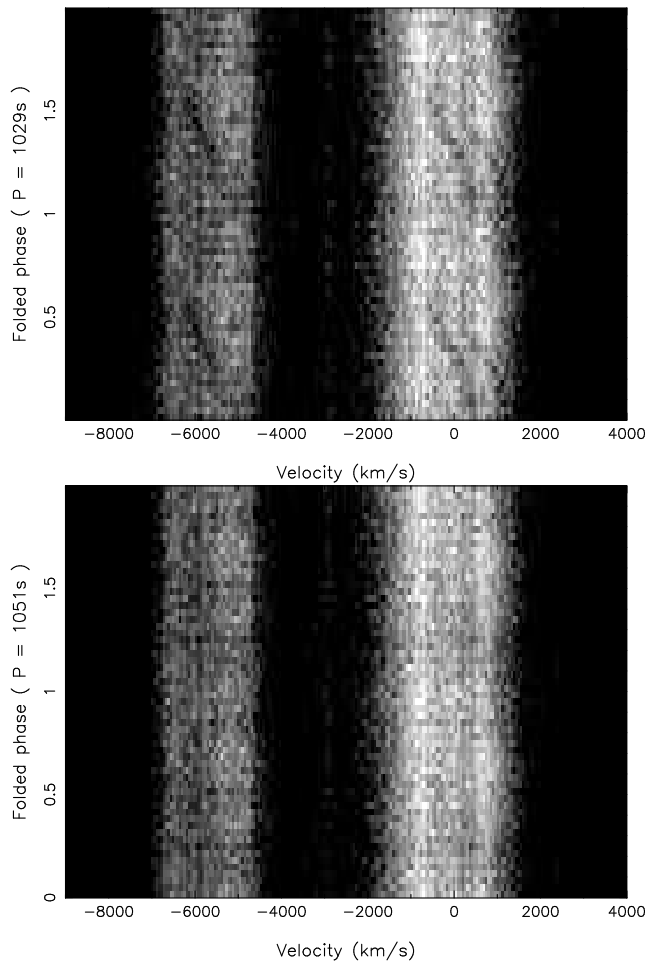

\psfig{figure=figure3a.ps,angle=-90,width=\columnwidth}
\psfig{figure=figure3b.ps,angle=-90,width=\columnwidth}
\caption[]{Trailed spectrogram of the He\,{\sc i} 4387 and 4471~\AA\
  line profiles (absorption in white). {\bf Top:} after folding on the
  candidate orbital period of 1028.73~s. {\bf Bottom:} after folding
  on 1051.2~s. The gray scale is chosen in order to highlight the weak
  emission components in the core of the lines. A clear emission
  component is visible in both lines when the data are folded on
  1029~s only, indicating that this is indeed the true orbital period
  of AM CVn.}
\label{folds}
\end{figure}

An emission component is seen to be moving periodically through the
line profiles when the spectra are folded on 1029~s, but not when they
are folded on 1051~s (Figure \ref{folds}). This S-wave is a
characteristic signature in the line profiles of accreting binaries
caused by the impact of the gas stream onto the accretion disc around
the primary \citep[e.g.][]{mar90,sr98,mar99}.  The phase resolved
spectroscopy thus provides a definite spectroscopic signature of the
true orbital period of AM CVn, and firmly establishes its binary
nature.

The S-wave is most clearly visible in the He\,{\sc i} lines at 4387
and 4471~\AA, but can also be discerned in the line at 4143~\AA.  The
He\,{\sc ii} lines at 4686 and 4199~\AA, however, do not show any
evidence for S-waves on the orbital period.  The S-wave is strongest
moving from redshift to blueshift, but only barely present when moving
from blue to red. This reflects the fact that the hotspot is mainly
visible when it is at the front of the system as observed from Earth.
Note that when the spot comes to the front the velocity of the stream
and of the disk at the impact point are directed away from the
observer and thus redshifted.

\subsection{Doppler tomography}\label{doppler}

As a next step we applied Doppler tomography in order to reveal the
velocity structure in the lines as well as to establish the exact
velocity of the S-wave.  Doppler mapping \citep{mh88} uses the time
dependent line profile shapes to reconstruct the distribution of line
emission/absorption in the corotating frame of the binary. It is a
widely applied tool in the study of the strong emission lines in CVs.
It has revealed a variety of structures associated with the accretion
disc around the primary in interacting binaries such as bright spots,
spiral arms and magnetic streams \citep[see for a review][]{mar01}.

\begin{figure}
\psfig{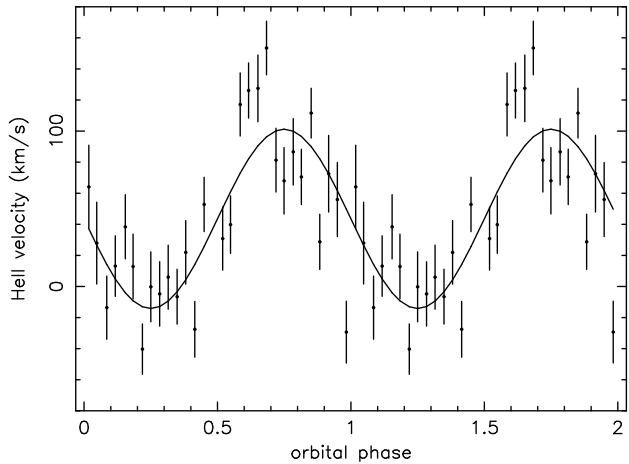}
\caption[]{The radial velocities of the He\,{\sc ii} line, derived
  from a Gaussian fit as a function of orbital phase. A sinusoidal fit
  is used to establish the absolute orbital phases and derive an upper
  limit to the radial velocity of the primary. }
\label{he2fit}
\end{figure}

So far we have used an arbitrary zero point for our orbital phases
since the absolute phase is not known from the photometry.  In order
to estimate the absolute orbital phase of the binary, where phase 0.0
is defined as superior conjunction of the primary star, we analysed
the line profile behaviour when folded on the 1029~s orbital period.
The line profiles are complex with a mix of absorption and emission
components present at any given time. The He\,{\sc ii} 4686~\AA\ line
is an exception and consist of a single emission component
superimposed on an absorption trough, and shows no evidence for an
S-wave.  We fitted the He\,{\sc ii} emission component with a simple
Gaussian, in order to measure its radial velocity as a function of
binary phase.  The radial velocity of the He\,{\sc ii} line shows a
systematic variation with an amplitude of $53 \pm 6 $km~s$^{-1}$ when
fitted with a simple sine-function (Fig.~\ref{he2fit}). If the
emission is associated with the accretion flow around the primary,
this gives us an indication of the motion of the primary white dwarf.
Care must be taken since asymmetries in the distribution of the
He\,{\sc ii} emission around the primary also introduces apparent
radial velocity shifts. So rather than assuming that the derived
velocity is indeed the projected velocity of the white dwarf ($K_1$),
we merely use the relative phasing in order to construct a more
reliable zero point for our phases.  The derived velocity amplitude
can be taken as an upper limit to the projected velocity of the white
dwarf.  Using the convention that orbital phase zero corresponds to
superior conjunction of the primary, we then derive the following
orbital ephemeris for AM CVn;

\[
T_0 (HJD)  = 2450140.6135(2) + 0.011906623(3) E 
\]

\noindent with the formal uncertainty of the zero point indicated between 
brackets and the period taken from \citet{spk+99}. In all plots,
the orbital phases shown are the result of folding using the above
ephemeris.

\begin{figure}
\psfig{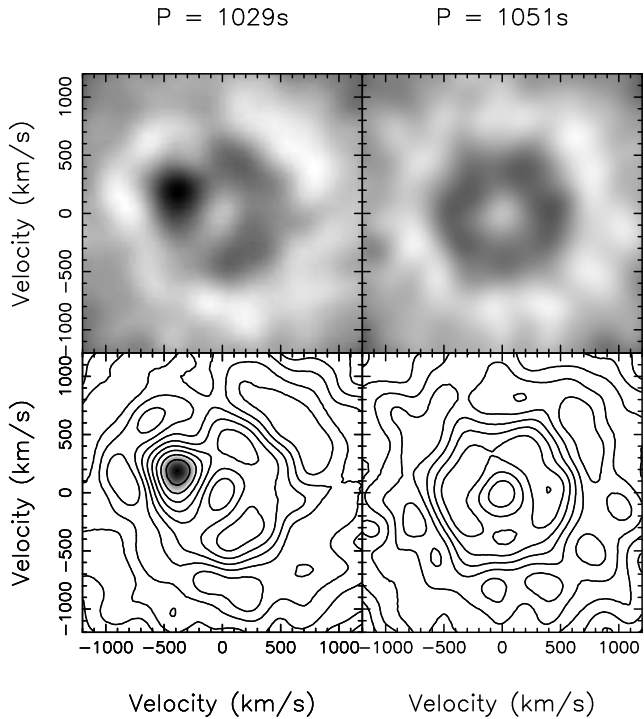}
\caption[]{
  Filtered back-projection of the He\,{\sc i} 4471~\AA\ line profiles
  on the orbital period of 1029~s as well as the photometric period of
  1051~s. The hotspot produces a prominent spot superposed on weak
  disc emission if folded on the correct period only. (Darker
  indicates more emission). The bottom two panels are the same images
  displayed as a contour plot. Any emission significantly brighter
  than the emission from the disc ring is overplotted as a gray scale,
  revealing the strong spot in the left tomogram. }
\label{backpro}
\end{figure}

There are two methods of calculating Doppler tomograms from phase
resolved spectroscopy. One is a straightforward back projection in
conjunction with a Fourier filter \citep{hor91}, the other applies
maximum entropy regularisation \citep{mh88}. Our goal is to isolate
the weak emission components in the cores of the He\,{\sc i}
absorption lines.  To this end the filtered back projection method is
preferred, since a maximum entropy reconstruction requires pure
emission line profiles.  Although more prone to reconstruction
artifacts, the back-projection method can be applied to the complex
line profiles of AM CVn without additional data processing.  The
results of back projecting the line profiles of the He\,{\sc i}
4471~\AA\ line are plotted in Fig.~\ref{backpro}.  Again we compare
the back projection of the line profiles folded on the two periods;
1029~s and 1051~s. In the case of 1029~s, the S-wave maps into a
prominent spot in the Doppler tomogram that is absent when
folded on 1051~s. The location of the emission spot is where hotspot
emission is expected, giving us confidence in the orbital
ephemeris that is derived and our previous interpretation of the
S-wave as due to the hotspot on the outer edge of the accretion disc.

\subsection{The radial velocity of the mass donor}\label{K2}

Apart from the hotspot itself, a weak ring of emission is visible in
Fig.~\ref{backpro} from the rest of the accretion disc. At higher
velocities this weak emission turns into absorption which produces the
broad deep absorption wings in the lines.  Having established the true
orbital period of AM CVn, we can also constrain other system
parameters using the position of the bright spot.  The hotspot is the
result of local heating and dissipation at the outer rim of the disc
where the infalling gas stream impacts.  In most cases, the emission
has the velocity of the free falling stream \citep[e.g.][]{mh90}.
However, in other cases the velocity of the hot spot gas appears to be
a mix of the fall velocity as well as the local velocity flow in the
disc at the impact point \citep{mar90}.  This unfortunately means that
a straightforward fit to the position of the hotspot using single
particle trajectories is unreliable.  However, some interesting limits
can still be obtained.  First of all the velocity amplitude of the
spot itself ($403 \pm 15$ km s$^{-1}$) provides an upper limit to the
possible projected velocity of the mass donor star ($K_2$), since the
velocity of its centre of mass cannot be larger than that of the
hotspot itself. Very low values of $K_2$ also lead to gas stream
trajectories that are incompatible with the data. For a given choice
of $K_1,K_2$, we can calculate the trajectory of the gas stream as
well as the velocity vector of the disc flow in order to isolate those
parts in the tomogram where hot spot emission is possible. We shall
see later that the mass ratio $q (=M_2/M_1)$ of AM CVn is most likely
0.087, which means that for a given choice of $K_2$, $K_1$ is derived
from $K_1 = q K_2$.

\begin{figure}
\psfig{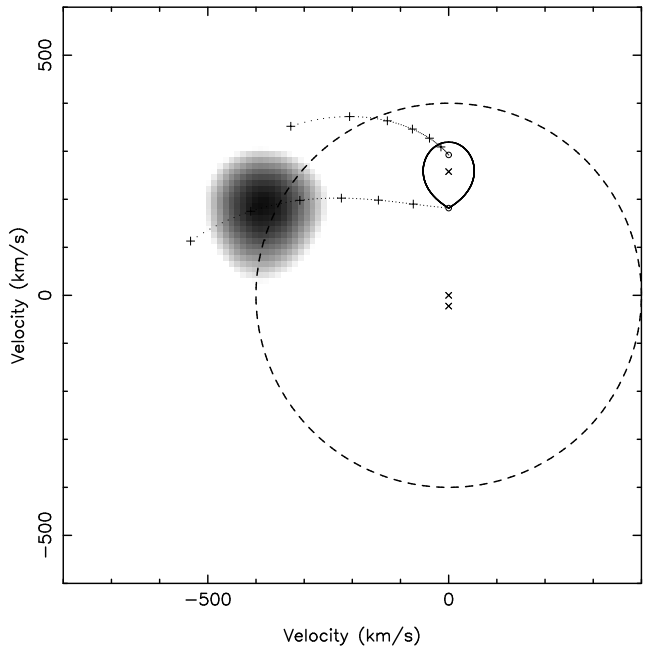}
\caption[]{Ballistic trajectories that fit to the location of the
  hotspot. The observed position of the hotspot is plotted as a gray
  scale.  Overplotted is the best fit ballistic stream trajectory, if
  the hotspot corresponds to pure free fall velocities. The
  corresponding velocities of the disc are indicated by the second
  trajectory above the ballistic one. The observed hotspot velocities
  need to lie in between these two trajectories for a given choice of
  $K_2$, leading to possible $K_2$ values between 210 and 265 km
  s$^{-1}$.  The Roche lobed shape indicates the location of the mass
  donor star in the tomogram and the three crosses the centres
    of mass of the two components and the binary system.}
\label{stream}
\end{figure}

If we assume that the hotspot velocities in AM CVn reflect the
velocity of the ballistic gas stream, we can achieve a good fit to the
observed tomogram using $K_2 = 260$ km s$^{-1}$ and $K_1 = 23$ km s$^{-1}$
(Fig.~\ref{stream}). If we relax our assumptions and only require that
the observed velocities of the hotspot lie anywhere between the
trajectory of the gas stream, and that of the disc, we can constrain
$K_2$ to lie between 210 and 265 km s$^{-1}$. The assumed value of the
mass ratio has only a small effect on the derived values for $K_2$.

The least constrained range for $K_2$ is obtained if we also consider
the zero point of our orbital phases to be a free parameter. Then, in
principle, $K_2$ can lie between 200 and 400 km s$^{-1}$. Note,
however, that in order to fit the data with $K_2$ $>$ 280 km s$^{-1}$,
we need to apply an arbitrary phase shift in order for the stream to
cross the hotspot, which also leads to an unrealistic impact radius
that is very close to the white dwarf. We can thus firmly constrain
the value of $K_2$ to lie between 210 and 280 km s$^{-1}$.

\subsection{Disc precession and superhumps}\label{precession}

\begin{figure}
\psfig{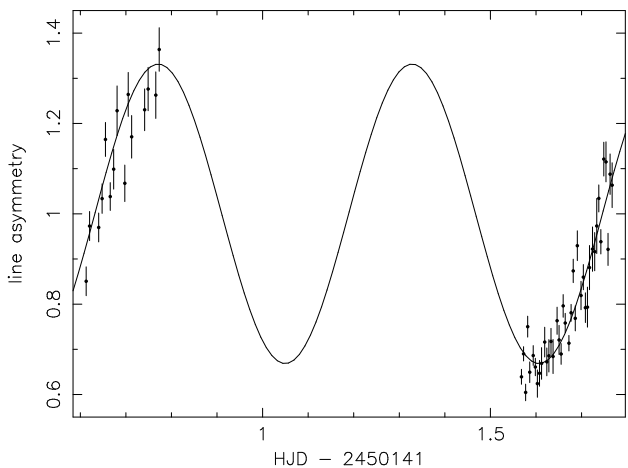}
\caption[]{The change in relative absorption on the redshifted versus
  the blueshifted side across the 13.38 hour precession cycle.
  Plotted is the line asymmetry derived from a double Gaussian fit as
  a function of HJD. A least squares sinusoidal fit is overplotted. }
\label{skewness}
\end{figure}

\citet{phs93} have found a 13.38~hr period in the skewness of the
He\,{\sc i} absorption lines. From Fig.~\ref{fig:spectra} we see
indeed that the shape of the absorption lines changes during the 5 hrs
of the total observation.  This can be interpreted as the varying
double peaked absorption profiles of a precessing accretion disc
\citep{phs93}. We measured the skewness of the line profiles
in our data by fitting multiple Gaussians to the absorption lines as
follows.  Since there is evidence for emission in the line cores,
which modulates on the much shorter orbital period compared to the
slow changes in the absorption lines on the precessing periods, the
central emission was masked.  Double Gaussians were fitted to the
absorption components in order to measure the relative strength of the
blue shifted absorption relative to the redshifted absorption. The
spectra of both nights were grouped into bins of 10 consecutive
spectra to improve signal to noise and the line asymmetry was measured
accordingly. This double Gaussian method was found to be more reliable
than calculating the formal skewness of the line as a whole since the
emission components can be masked so as not to distort the
measurements. Fig.~\ref{skewness} plots the derived line asymmetry
together with a sinusoidal least squares fit.  We do not cover the
full precession cycle, so cannot improve on the precession period
derived by Patterson et al. (13.38 hours) and fixed the period in the
fit.  The asymmetry is expressed as the ratio between the depths of
both Gaussians.  The relative contribution of the red versus blue
shifted side of the disc thus changes with an amplitude of 33\% across
the precessing cycle. Adopting the very simple picture
  sketched in \citet{phs93} using their $\phi_0$ as phase, we coved
  phases 0.3 -- 0.75 on the first night and phases 0.2 -- 0.6 on the
  second night.

\section{Discussion}\label{discussion}

The fact that AM CVn behaves very similarly to the hydrogen rich
superhump systems \citep{phs93} suggests that the superhump phenomena
has the same origin \citep{war95}. It could be explained by
  precession of an eccentric disk, which is caused by a tidal
  resonance between the disc and the companion which is found for mass
  transferring systems with a mass ratio smaller than 0.22
  \cite{whi88} up to 0.33 \citep{mur00}. \citet{sw98} find a more
  complex behaviour of the disk, including disk shape changes and
  spiral shocks in the disk, but a similar `precession' period.  The
numerical simulations of \citet{ho90,ho93} show that there is a
relation between the precession period, the orbital period and the
mass ratio \citep{war95}
\begin{equation}
\frac{P_{\rm prec}}{P_{\rm orb}} = A \frac{1 + q}{q}
\end{equation} 
where $A \approx 3.73$ for systems with $q < 0.1$. Because of the
tidal origin of the phenomenon there is no reason to expect a
different relation for helium discs. Using 1028.73~s for the orbital
period and 13.38~hr for the precession period, this relation gives a
mass ratio for AM CVn of 0.087.

From the well known fact that for a Roche lobe filling star the period
only depends on the mass and radius of the donor we can calculate the
mass of the donor in AM CVn from the period and the mass -- radius
relation. From the two evolutionary scenarios that are proposed to
lead to AM CVn stars, two mass -- radius relations for the donor stars
are found \citep{war95,npv+00}. From the mass ratio the companion mass
is found. In Table~\ref{tab:parameters} we give the two solutions for
AM CVn using the mass -- radius relation as given in \citet{npv+00}.

From the system parameters we can calculate the radial velocities of
the two components, which we can compare to the values of $K_2$ as
derived from the Doppler tomograms. In Table~\ref{tab:parameters} we
give the expected values of the radial velocities ($K_{1,2}/\sin i$)
and the resulting values for the inclination assuming $K_2 = 260$
km~s$^{-1}$.

%
%
%
%
\begin{table}
\caption[]{System parameters for the two different mass radius
  relations: deg. for fully degenerate mass -- radius relation for white
  dwarfs and semi for the semi-degenerate mass radius relation for the 
  final products of helium CVs \citep[from][see
  \citealt{npv+00}]{tf89}. The inclination is determined assuming $K_2$ =
  260 km~s$^{-1}$.
}
\label{tab:parameters}
\begin{center}
\begin{tabular}{lrrrrrr} \hline \hline
 & $M_2$ & $M_1$ & K1/$\sin i$ & K2/$\sin i$ & $i$ \\ 
 & (\msun) & (\msun) &  (km~s$^{-1}$) &  (km~s$^{-1}$) & $^\circ$ \\ \hline 
deg. & 0.033 & 0.38 & 55.5 & 637.4 & 24.1 \\
semi & 0.114 & 1.31 & 83.7 & 961.8 & 15.7 \\ \hline \hline
\end{tabular}
\end{center}
\end{table}

The derived inclinations are low, consistent with the fact that no
eclipses are observed (which means $i < 78^\circ$ for $q = 0.087$).
The only clue to discern the two mass -- radius relations (and thus
the formation scenario) may be the fact that the hotspot is obscured
when moving from blue to red, implying not too low inclination. This
would slightly favour the fully degenerate mass radius relation.
However the mass ratio derived from the superhump period is
  too uncertain to draw firm conclusions.

The value of $K_1$ derived from the mass ratio of 0.087 (23 km
s$^{-1}$) seems incompatible with the interpretation of the radial
velocity of the He\,{\sc ii} line as being caused by the motion of the
primary. However it could be that the mass ratio determined from the
superhump period is incorrect for AM CVn. For example for a mass ratio
of 0.2, the value of $K_1$ would be 52 km s$^{-1}$. This also would
bring the mass of the primary down, which rules out the fully
degenerate mass -- radius relation, but gives somewhat more realistic
value for the primary mass in the case of the semi-degenerate mass --
radius relation \citep[see for more details][]{npv+00}.


Evidence for time variability of the strength of the hot spot was
found by folding both nights separately. Using the the first night
only, no discernible S-wave could de identified. However, the second
night alone, resulted in a clearly visible S-wave in several lines.
Maybe the hot spot strength varies with the precession period,
  allthough the precession phase shift between the two nights is
  small. The poor quality of the data of the first night makes it for
  the moment impossible to draw conclusions. Better sampling of the
  precession period offers opportunities to study the strength and
  position of the hot spot as a function of the precession cycle,
  allowing a better determination of the hot spot position and thus
  the mass ratio provided adequate signal to noise can be obtained.

\section{Conclusion}\label{conclusion}

We found a clear S-wave in spectra of AM CVn when folded on a period
of 1029~s, confirming the binary nature of AM CVn beyond any possible
remaining doubt and establishing 1028.73~s as the true orbital period.

The dominant photometrical period of 1051.2~s can be interpreted as
the beat between the orbital period and the 13.38~hr period that has
been found in the skewness of the absorption lines, making the system
the helium equivalent of the permanent superhump systems. The 13.38 hr
period then is interpreted as the precession period of the accretion
disc. The shape of the absorption lines in our data set changes on the
same period.

By applying Doppler tomography we were able to constrain the value of
$K_2$ between 210 and 265 km s$^{-1}$, which given the formation
channels implies a low inclination (between $\sim15^\circ$ and
$\sim25^\circ$).

Our results open a new field of studying the details of these
intriguing ultra-compact binaries by phase resolved spectroscopy and
Doppler tomography.

\section*{Acknowledgments}
We thank Tom Marsh for motivating discussion and the use of the MOLLY
data reduction package and the referee M. Wood for valuable
comments. The William Herschel Telescope is operated on the island of
La Palma by the Isaac Newton Group in the Spanish Observatorio del
Roque de los Muchachos of the Instituto de Astrofisica de Canaria.
This paper makes us of data obtained from the Isaac Newton Group
Archive which is maintained as part of the Astronomical Data Centre at
the Institute of Astronomy, Cambridge. GN is supported by NWO Spinoza
grant 08-0 to E.~P.~J.~van den Heuvel, DS is supported by a PPARC
Fellowship and PJG is supported by a CfA fellowship.

\bibliography{journals,binaries}
\bibliographystyle{NBmn}

\label{lastpage}

\end{document}